\documentclass[aps,prb,twocolumn,superscriptaddress,showpacs]{revtex4}
\usepackage{amsfonts}
\usepackage{amssymb}
\usepackage{amsmath}
\usepackage{graphicx}

\usepackage{dcolumn}   
\usepackage{bm}        
\usepackage{flafter}

\begin{document}


\title{Observation of the in-plane magnetic field-induced phase transitions in FeSe}

\date{\today}
\author{Jong Mok Ok}
\affiliation{Department of Physics, Pohang University of Science and Technology, Pohang 37673, Korea}
\affiliation{Center for Artificial Low Dimensional Electronic Systems, Institute for Basic Science, Pohang 37673, Republic of Korea}
\author{Chang Il Kwon}
\affiliation{Department of Physics, Pohang University of Science and Technology, Pohang 37673, Korea}
\affiliation{Center for Artificial Low Dimensional Electronic Systems, Institute for Basic Science, Pohang 37673, Republic of Korea}
\author{Yoshimitsu Kohama}
\affiliation{ISSP, University of Tokyo, Kashiwa, Chiba 277-8581, Japan}
\author{Jung Sang You}
\affiliation{Department of Physics, Pohang University of Science and Technology, Pohang 37673, Korea}
\affiliation{Center for Artificial Low Dimensional Electronic Systems, Institute for Basic Science, Pohang 37673, Republic of Korea}
\author{Sun Kyu Park}
\affiliation{Department of Physics, Pohang University of Science and Technology, Pohang 37673, Korea}
\affiliation{Center for Artificial Low Dimensional Electronic Systems, Institute for Basic Science, Pohang 37673, Republic of Korea}
\author{Ji-hye Kim}
\affiliation{Department of Physics, Kyungpook National University, Daegu 702-701, Republic of Korea}
\author{Y. J. Jo}
\affiliation{Department of Physics, Kyungpook National University, Daegu 702-701, Republic of Korea}
\author{E. S. Choi}
\affiliation{National High Magnetic Field Laboratory, Florida State University, Tallahassee, Florida 32310, USA}
\author{Koichi Kindo}
\affiliation{ISSP, University of Tokyo, Kashiwa, Chiba 277-8581, Japan}
\author{Woun Kang}
\affiliation{Department of Physics, Ewha Womans University, Seoul 03760, Republic of Korea}
\author{Ki Seok Kim}
\affiliation{Department of Physics, Pohang University of Science and Technology, Pohang 37673, Korea}
\author{E. G. Moon}
\affiliation{Department of Physics, Korea Academy Institute of Science and Technology, Daejeon 34141, Republic of Korea}
\author{A. Gurevich}
\affiliation{Department of Physics, Old Dominion University, Norfolk, VA 23529, USA}
\author{Jun Sung Kim}
\email{js.kim@postech.ac.kr}
\affiliation{Department of Physics, Pohang University of Science and Technology, Pohang 37673, Korea}
\affiliation{Center for Artificial Low Dimensional Electronic Systems, Institute for Basic Science, Pohang 37673, Republic of Korea}

\begin{abstract}
We investigate thermodynamic properties of FeSe under the in-plane magnetic fields using torque magnetometry, specific heat, magnetocaloric measurements. Below the upper critical field $H_{c2}$, we observed the field-induced anomalies at $H_1$ $\sim$ 15 T and $H_2$ $\sim$ 22 T near $H \parallel ab$ and below a characteristic temperature $T^*$ $\sim$ 2 K. The transition magnetic fields $H_1$ and $H_2$ exhibit negligible dependence on both temperature and field orientation. This contrasts to the strong temperature and angle dependence of $H_{c2}$, suggesting that these anomalies are attributed to the field-induced phase transitions, originating from the inherent spin-density-wave instability of quasipaticles near the superconducting gap minima or possible Flude-Ferrell-Larkin-Ovchinnikov state in the highly spin-polarized Fermi surfaces. Our observations imply that FeSe, an atypical multiband superconductor with extremely small Fermi energies, represents a unique model system for stabilizing unusual superconducting orders beyond the Pauli limit.
\end{abstract}
\smallskip


\maketitle

\section{INTRODUCTION}

In conventional superconductors 
spin polarization can destroy superconductivity when Zeeman energy surpasses the binding energy of Cooper pairs, known as the Pauli paramagnetic limit. In unconventional superconductors, however, exotic superconducting phases have been observed beyond the Pauli limit~\cite{FF,LO,ffloreview,organicreview,offlo4,hfq1,kenzelmann1,kenzelmann2,QSDW,kenzelmannreview}, often coexisting with complex magnetic orders. For example,
the Fulde-Ferrel-Larkin-Ovchinnikov (FFLO) state was proposed a half century ago,~\cite{FF,LO} in which $s$-wave Cooper pairs acquire a finite center-of-mass momentum to form a spatially modulated superconducting order.~\cite{ffloreview,organicreview,offlo4} Alternatively, spin-triplet pairing of field-induced quasiparticles can occur particularly in unconventional $d$-wave superconductors, inducing novel phases with mixture of spin-singlet and spin-triplet pair density waves.~\cite{hfq1,kenzelmann1,kenzelmann2,QSDW,kenzelmannreview} Growing evidence has revealed that the former and the latter are realized in highly anisotropic organic superconductors and a heavy fermion superconductor CeCoIn$_5$, respectively.~\cite{kenzelmannreview}
In both systems, however, 
the typical field-induced spin imbalance $\sigma$ = ($n_\uparrow$-$n_\downarrow$)/($n_\uparrow$+$n_\downarrow$), where $\uparrow$ and $\downarrow$ refer to the two spin components with densities $n_{\uparrow,\downarrow}$, is at most a few \%. 
Realization of the highly spin-imbalanced superconductivity and possible field-induced exotic phases have 
been remained elusive so far. 

Usually in single-band superconductors, Cooper pairs are broken by the orbital pair breaking effect or by the Pauli paramagnetic effect ~\cite{WHH}. The orbital pair breaking effect destroys superconductivity at the relatively low upper critical fields $H_{c2}$ with negligible spin imbalance. Even the orbital effect is largely suppressed, the Pauli paramagnetic effect limits the spin imbalance, typically $\sigma$ $\sim$ 10$^{-2}$. This is because the maximum Zeeman energy $\mu_B H_{c2}$ ($\mu_B$, Bohr magneton) is set by the superconducting gap $\Delta_{\rm SC}$ and usually much smaller than the Fermi energy $E_{\rm F}$. Multiband superconductors however may host the strongly spin imbalanced state. If the $E_{\rm F}$ is small in one band with a small $\Delta_{\rm SC}$ and the superconductivity is maintained by the other band with a relatively larger $\Delta_{\rm SC}$ ~\cite{review1,review2}, spin polarization can be significantly enhanced under high magnetic fields .

Such a candidate is FeSe with $\sigma$ $\sim$ 10$^{-1}$. FeSe is
known to be in the so-called BCS-BEC (Bose-Einstein condensate) crossover regime ($\Delta_{SC} \sim E_F$)~\cite{FeSe:BCS,FeSe:BCS2}, and also in
the Pauli limiting regime ($\mu_B H_{c2} \sim \Delta_{SC}$)~\cite{FeSe:BCS,FeSeHc2_1} under in-plane magnetic fields.
Possible magnetic field induced phase transitions have been suggested under the out-of-plane fields~\cite{FeSe:BCS}, and very recently, also in the in-plane fields~\cite{FeSe_matsuda}. Therefore FeSe can be a model system to study whether, and if so how, competing magnetic or superconducting instabilities trigger exotic field-induced phases in multiband superconductors~\cite{FFLO0,FFLO1,FFLO2} or in the large spin imbalance regime~\cite{spinim1,spinim2,spinim3}.
In this work, using torque magnetometry, specific heat, and magnetocaloric measurements, we identified successive anomalies with increasing in-plane magnetic field, below $H_{c2}$ at low temperatures and in the clean limit. These results evidence that unusual superconducting phases are indeed stabilized in FeSe at high in-plane magnetic fields.

\section{Experiments}
Single crystals of FeSe were grown by a KCl-AlCl$_3$ flux technique~\cite{FeSe:growth,FeSeNMR}. The resistivity of single crystals
were measured using the standard four-probe method in a Physical Property Measurement
System (PPMS-14T, Quantum Design) in order to determine the residual resistivity ratio
(RRR). Single crystals with the RRR value higher than 35 were used in the high magnetic
field experiments unless otherwise noted (Supplementary Fig. S1).

 The upper critical field $H_{c2}$ of FeSe was determined using the resistivity and the tunnel diode oscillator (TDO) frequency measurements at high magnetic fields. We measured the resistive transition under magnetic field up to 30 T in National High Magnetic Field Laboratory (NHMFL), Tallahassee. For resistivity measurements, a room temperature cured silver paste/epoxy was used to connect gold wires to the sample. The TDO measurements up to 30 T were performed in International MegaGauss Science Laboratory in Institute for Solid State Physics (ISSP), University of Tokyo. In the TDO measurements, we recorded the change of the radio frequency in an LC circuit with a crystal wound in a copper coil. For pressure experiments, quasi-hydrostatic pressure was applied up to 35 kbar using a home-made indenter-type pressure cell. The superconducting transition of lead, mounted next to the FeSe crystal, was used to determine pressure inside the cell at low temperatures (Supplementary Fig. S3).

 The thermodynamic anomalies in the superconducting state were observed using the measurements on torque magnetometry, magnetocaloric effect, and specific heat under magnetic fields up to 30 T at NHMFL. For torque magnetometry measurements, we used a miniature Seiko piezoresistive cantilever. On top of the cantilever we mounted a small single crystal, typically 100$\times$100$\times$20 $\mu$m$^3$, extracted from the crystal whose RRR value was determined from the preceding resistivity measurements. During the up- and down-sweeps of the applied magnetic field, the resistance of the piezoelectric cantilever was monitored, which is proportional to the magnetic torque acting on the cantilever. The in-plane magnetic field is applied along the sample edge (Supplementary Fig. S5), which is nearly along the diagonal direction of the orthorhombic lattice. For magneto-caloric effect and specific heat measurements, the samples were mounted on top of a bare chip Cernox or RuO$_2$ thermometer, suspended by manganin wires (diameter $\sim$ 10 $\mu$m) as shown in Supplementary Fig. S6. Detailed information on experiments are available in Supplementary Materials.

\section{Results}
\subsection{Characteristics of FeSe}

\begin{figure}[t]
\centering
\includegraphics[width=8cm, bb=36 60 340 360]{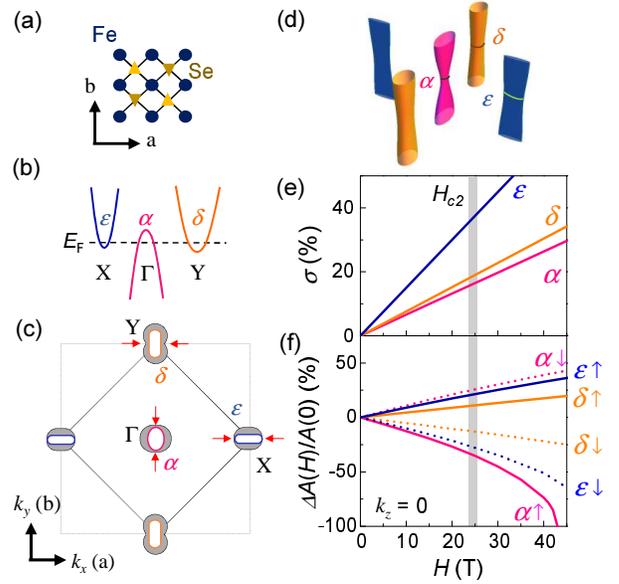}
\caption{\label{fig2} (a) Schematic crystal structure of FeSe. Fe atoms form an almost square lattice in the plane, and Se atoms are located above and below the center of Fe$_4$ plaquettes. The orthorhombic crystal axes, $a$ and $b$ are indicated by the arrows. (b) Schematic band structure of FeSe. One hole band is centered at the $\rm{\Gamma}$ point and denoted as $\alpha$, while two electron bands are located at Y and X points and denoted as $\delta$ and $\varepsilon$, respectively. The Fermi level $E_\textrm{F}$ is indicated by the dashed line. (c) Fermi surfaces (FSs) of FeSe in the $k_z$ = 0 plane. The gray shades display the anisotropic superconducting gap of each FS. The superconducting gap minima are marked by red arrows. (d) Fermi surfaces of FeSe (e,f) Spin imbalance $\sigma$ (e) and the change of the cross-sectional area $\Delta A(H)$ in the $k_z$ = 0 plane (f), as a function of magnetic field. For the hole FS ($\alpha$), $A(H)$ of $\downarrow$ spins become larger with increasing magnetic field, whereas $A(H)$ of $\uparrow$ spins is reduced. The electron FSs ($\delta$ and $\epsilon$) show the opposite field dependence. The in-plane upper critical field ($H_{c2}$) is indicated by a grey vertical line.
}
\end{figure}
FeSe is the simplest but an atypical member of FeSCs, which has $T_c$ $\approx$ 9 K and is composed of edge-shared FeSe$_4$ tetrahedra layers in van der Waals (vdW) structure~\cite{FeSereview}. The observed in-plane upper critical field $H_{c2}^{ab} \approx$ 25 T is well above the Pauli limiting field $H_P \approx$ 15.6 T at the BCS limit ($H_P$= 1.84$T_c$), which indicates that FeSe is in the Pauli limit under in-plane magnetic fields ($\mu_B H_{c2} \sim \Delta_{SC}$). In addition, high-quality FeSe single crystals with a high residual resistive ratio (RRR) $\geq$ 35 are in the clean limit, as confirmed by magnetic quantum oscillation studies~\cite{FeSe:QO1,FeSe:QO2,FeSe:QO3} (Supplementary Fig. S1). More importantly, FeSe has exceptionally small Fermi energies~\cite{FeSe:BCS,FeSe:BCS2}. In FeSe, three Fermi surfaces (FSs) exist; one hole FS is centered at the $\Gamma$ point and denoted as $\alpha$, two electron FSs are located at the X and Y points and denoted as $\delta$ and $\epsilon$, respectively (Figs. 1b and 1c)~\cite{FeSe:QO1,FeSe:QO2,FeSe:QO3,FeSe:QPI,FeSe:ARPES}. The low energy structure of each FS can be described by $H=\hbar^2 k_x^2/2m_x+\hbar^2 k_y^2/2m_y+2t\cos(k_z)$, where $m_{x,y}$ is the effective mass along the in-plane momentum $k_{x,y}$ directions, and $t$ is the interlayer hopping parameter. We note that this model does not capture the detailed FS shapes, such as a bow-tie shape of $\epsilon$ FS, but provide a qualitative picture of FSs under high magnetic fields. Here the $x$ axis ($y$ axis) is defined to be parallel to the $a$ axis ($b$ axis) of the Fe orthorhombic lattice (Fig. 1a). Based on recent quantum oscillations and angle-resolved photoemission studies~\cite{FeSe:QO1,FeSe:QO2,FeSe:QO3,FeSe:ARPES}, the parameters for the low energy structures can be obtained as summarized in Supplementary Table S1. The Fermi energies for these bands are $E_F \approx$ 7.9 meV, 6.1 meV, and 2.9 meV for $\alpha$, $\delta$ and $\epsilon$ pockets, respectively. These values are comparable with their superconducting energies, $\Delta_{\rm SC}$= 1.64 meV, 1.53 meV, and 0.39 meV for $\alpha$, $\delta$ and $\epsilon$ pockets, respectively~\cite{FeSe:QPI,FeSe:Cp}. The ratio $\Delta_{SC}/E_F$ is $\sim$ 0.2 for all the FSs, confirming that FeSe is in the BCS-BEC crossover regime.

Under magnetic fields, the Zeeman effect induces imbalance of spin population as well as momentum mismatch between FSs of $\uparrow$ and $\downarrow$ spins. In FeSe with three FSs, spin imbalance and momentum mismatch are different at each FS, as shown in Figs. 1e and 1f. The spin imbalance in three FSs increases with magnetic fields and reaches up to ~15\% for $\alpha$ and $\delta$ FSs and ~40\% for $\epsilon$ at $H_{c2}^{ab} \approx$ 25 T s, assuming the $g$-factor of 2. These are much larger than the typical value of a few \% in other superconductors~\cite{FeSe:BCS}. The momentum mismatch is also significant and is quantified by the size difference of the FS, $\Delta A$ = $A_\uparrow$-$A_\downarrow$, where $A$ denotes the cross-sectional area of each FS in the $k_z$ planes. The ratio $\Delta A(H)/A(0)$ in the $k_z$=0 plane becomes 10\%-50\% at $H_{c2}^{ab} \approx$ 25 T and is particularly large for the $\alpha$-FS having a significant $k_z$ warping (Fig.1f and Supplementary Fig. S8). The $\alpha$-FS with $\uparrow$ spin in the $k_z$=0 plane may disappear at $\sim$ 40 T, resulting in a magnetic Lifshitz transition~\cite{ptok} well above the in-plane $H_{c2}^{ab}$.

\subsection{Upper critical fields}

\begin{figure}[ht]
\centering
\includegraphics[width=8cm, bb=30 60 310 420]{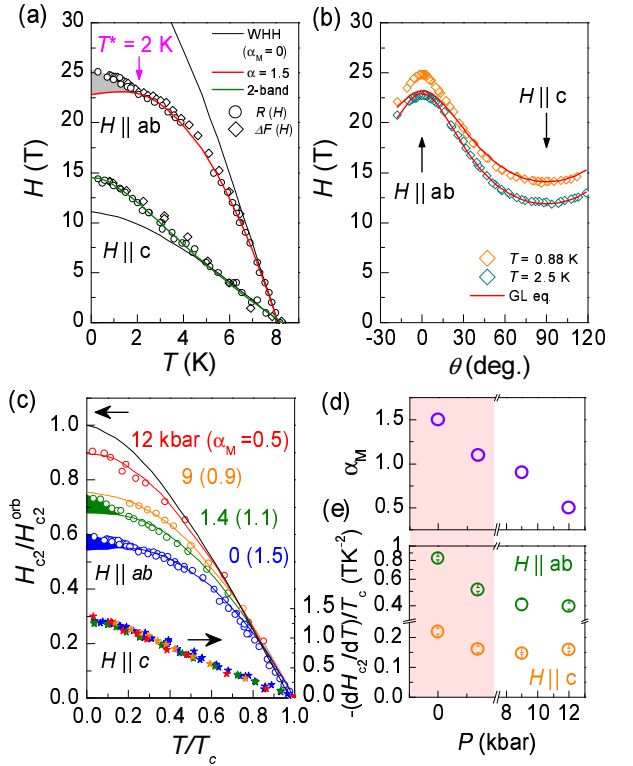}
\caption{\label{fig2} 
(a) Temperature dependent upper critical field $H_{c2}(T)$ of FeSe, estimated from resistivity (open circle) and TDO frequency measurements (open diamond). The WHH model predictions with $\alpha_{\rm M}$ = 0 (black line) and $\alpha_{\rm M}$ = 1.5 (red line) are presented for comparison. For $H \parallel ab$, strong enhancement of $H_{c2}^{ab}(T)$ below $T^*$ = 2 K, indicated by shaded area, cannot be explained by the WHH model. For $H \parallel c$, the two-band model curve (green line) well matches with the $H_{c2}^c(T)$ data.  (b) The field-angle dependence of $H_{c2}(\theta)$ at two different temperatures, $T$ = 0.88 K and 2.5 K, below and above $T^*$. The tilting angle ($\theta$) of the applied magnetic field is defined against the $ab$ plane. The anisotropic Ginzburg-Landau (GL) model (red line) reproduces nicely the $H_{c2}$ data taken at $T$ = 2.5 K ($> T^*$), while the clear enhancement of $H_{c2}$ is observed at $T$ = 0.88 K ($< T^*$) in the angle range of $|\theta| \lesssim 15^{\circ}$.
(c) Temperature dependence of $H_{c2}^{ab}(T)$ and $H_{c2}^c(T)$, taken at various external pressures and normalized by the orbital critical field $H_{c2}^{\rm orb}(0)$ = 0.69$T_c|dH_{c2}/dT|_{Tc}$. For $H \parallel ab$, the WHH model curves (solid lines) with different Maki parameters $\alpha_{\rm{M}}$, indicated by the numbers in parenthesis, agree well with the experimental data. With increasing pressure, $H_{c2}^{ab}(T)$ curves become closer to the orbital limiting case with $\alpha_{\rm{M}}$ = 0, and at the same time the enhancement of $H_{c2}$ is suppressed systematically. For $H \parallel c$, in contrast, all the normalized $H_{c2}(T)$ curves at different pressures overlap with each other. (d, e) Pressure dependence of Maki parameter $\alpha_{\rm M}$ (d) and the slope of $H_{c2}$ near $T_c$ for $H \parallel ab$, normalized by $T_c$ (e).}
\end{figure}

Having established that FeSe is simultaneously in the clean limit, in the BCS-BEC crossover regime, and in the Pauli limit under in-plane magnetic fields, we now discuss its magnetic phase diagram. Using the tunnel-diode oscillator frequency and the resistivity measurements, we obtained upper critical field $H_{c2}$ as a function of temperature (Fig. 2a) and field orientation (Fig. 2b), which are consistent with each other and independent of the types of measurements or criteria for determining $H_{c2}$ (Supplementary Fig. S2). For $H \parallel c$, the zero-temperature upper critical field $H_{c2}$(0) is estimated to be $H_{c2}^c \approx$ 15 T. At low temperatures, $H_{c2}^c (T)$ data deviate from the conventional Werthamer-Helfand-Hohenberg (WHH) prediction, which is well explained by the two-band model, as found in other FeSCs with significant interband interactions~\cite{review1,review2}. We note that the field-induced phase transition~\cite{FeSe:BCS} observed at $H$ $\sim$ 15 T for $H\parallel c$ does not affect the $H_{c2}^c(T)$ behavior.

In contrast, near $H \parallel ab$, we observe unusual behaviors of $H_{c2}$ in both the temperature- and the field angle-dependences. Here the in-plane magnetic field is applied along the diagonal direction of the orthogonal lattice (Fig. 1a) and thus denoted as $H\parallel ab$. In this case, the quasi-2D nature of FeSe suppresses the orbital pair breaking effect and enhances $H_{c2}^{ab}(0)$ up to $\sim$ 25 T, larger than $H_P$ (Fig. 2a). Usually in FeSCs, the Pauli-limiting effect is more important than the multiband orbital pair breaking effect under in-plane magnetic fields at low temperatures~\cite{review1,review2}. With a large Maki parameter $\alpha_M$=$\sqrt{2} H_{c2}^{orb}/H_P$ $>$ 1, where $H_{c2}^{orb}$ is the orbital-limiting upper critical field ($H_{c2}^{orb} =0.69T_c |dH_{c2}/dT|_{T_c}$), $H_{c2}^{ab}(T)$ curves of FeSCs, such as LiFeAs~\cite{Li111:FFLO}, KFe$_2$As$_2$~\cite{K122:FFLO}, exhibit a saturation behavior at low temperatures and are well described by the single-band WHH model~\cite{review1,review2}. In FeSe, however, we observed a weak but discernable kink in $H_{c2}^{ab}(T)$ curve at $T^*$ $\approx$ 2 K. Consistently the single-band WHH model with $\alpha_M$= 1.5 reproduces the measured $H_{c2}^{ab} (T)$ above $T^*$, but not its unusual upturn below $T^*\approx$ 2 K. Furthermore, at $T$= 0.88 K, just below $T^*$, $H_{c2}(\theta)$ as a function of field angle ($\theta$) clearly deviates from the GL model near $H\parallel ab$ with $\theta \leq 15^{\circ}$. 
This contrasts the $H_{c2}(\theta)$ data taken at $T$ = 2.5 K, just above the $T^*$, which follows nicely the anisotropic Ginzburg-Landau (GL) equation, $H_{c2}(\theta)=H_{c2}^c/\sqrt{(\cos^2\theta+\Gamma_H^{-2} \sin^2\theta})$, where $\Gamma_H$ is the mass anisotropy $m_{ab}/m_c$ and $\theta$ is the tilting angle of the field from the in-plane orientation. The deviations of $H_{c2}^{ab} (T)$ and $H_{c2}(\theta)$ from the WWH and GL models can in principle be related to the multiband effect~\cite{review1,review2}. This, however, cannot explain the kink of $H_{c2}^{ab} (T)$ at $T^*$ suggesting possible field-induced phase transitions, as discuss in the following section.

The pressure dependence of the low-temperature upturn in $H_{c2}^{ab}(T)$ suggests that the observed upturn in $H_{c2}^{ab} (T)$ is intimately related to the Pauli-limiting effect. It has been known that FS reconstruction occurs at a critical pressure $P_c$ $\sim$ 8 kbar~\cite{FeSe:QO3}, suggested by quantum oscillations and upper critical field studies~\cite{}. With external pressures below and above $P_c$, we found a systematic variation of $H_{c2}(T)$. For $H \parallel c$, $H_{c2}^{c} (T)$ curves, taken at different pressures and by $H_{c2}^{orb}$, follow the same temperature dependence as clearly shown in Fig. 2c. This implies that the relative ratios in the sizes or the diffusivity between dominant hole and electron pockets of FeSe are not significantly changed across $P_c$. On contrary, the Maki parameter $\alpha_M$, taken from the fit of $H_{c2}^{ab} (T)$ curves to the WHH model, is reduced systematically with pressure (Fig. 2d). At high pressures, therefore, the orbital-limiting effect becomes more important for determining $H_{c2}^{ab}$ than the Pauli limiting effect. This is consistent with the fact that the slops of $H_{c2}^{ab}$ at $T_c$, $i.e.$ $dH_{c2}^{ab}/dT|_{T_c}$ for $H\parallel ab$, that is proportional to the Fermi velocity along the $c$ axis, is reduced with increasing pressure (Fig. 2e)~\cite{Hc2slope}.
Concomitantly, the unusual upturn in $H_{c2}^{ab} (T)$ at low temperatures is systematically suppressed and eventually disappears for $\alpha_M$ $<$ 1 (Fig. 2c). These observations suggest the observed upturn in $H_{c2}^{ab} (T)$ is not simply due to the multiband effect, but may reflect formation of field-induced phase beyond the Pauli limit. In fact, this low temperature behavior in $H_{c2}^{ab} (T)$ resembles those of the unconventional Pauli-limiting superconductors, $\kappa$-(BEDT-TTF)$_2$Cu(NCS)$_2$ and CeCoIn$_5$ (Supplementary Fig. S2)~\cite{organicreview,kenzelmannreview}, and is consistent with the recent work on FeSe~\cite{FeSe_matsuda}.

\subsection{Magnetic field-induced phase transitions}
\begin{figure*}[ht]
\centering
\includegraphics[width=17cm, bb=30 60 570 260]{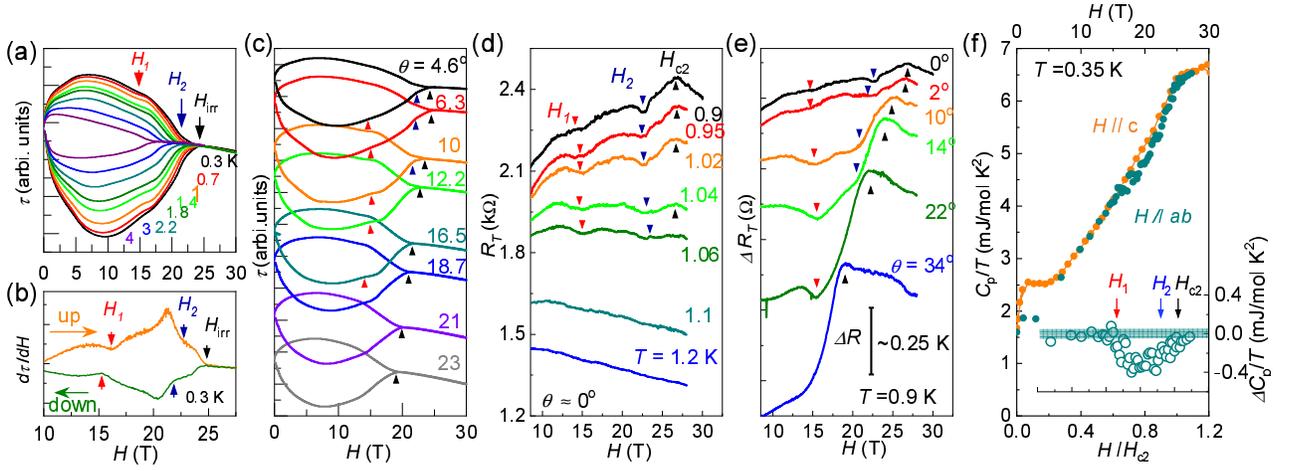}
\caption{\label{fig3} 
(a) Magnetic field dependent torque $\tau(H)$ close to $H\parallel ab$ at various temperatures. The irreversibility field at $H_{irr} \sim$ 25 T and anomalies at $H_1 \sim 15$ T and $H_2 \sim 22$ T are indicated by black, red and blue arrows, respectively. (b) The field-derivative curve $d\tau(H)/dH$ showing clear anomalies at $H_1$ and $H_2$. (c) Magnetic field dependent torque $\tau(H)$ at various field angles. 
(d) Magnetic field dependence of the magnetocaloric effect at different temperatures. The field dependence of the sensor resistance, monitoring the inverse of the sample temperature, exhibits clear anomalies at $H_1$, $H_2$, and $H_{irr}$. 
(e) Magnetic field dependence of the magnetocaloric effect at various tilting angles of the applied magnetic field. $H_1(\theta)$ and $H_2(\theta)$ from the magnetocaloric effect are nearly independent of the field angle $\theta$. 
The anomaly at $H_{irr}(\theta)$ becomes stronger with increasing $\theta$ due to the heating effect from vortex dynamics. (f) Magnetic field dependent specific heat $C_s/T$, taken at $T$ = 0.35 K, as a function of the normalized field $h = H/H_{c2}$ for $H\parallel ab$ and $H\parallel c$. The difference $\Delta C_s(h)/T$ = $C_s(h)/T|_{H\parallel ab}$-$C_s(h)/T|_{H\parallel c}$, is shown in the inset, together with the arrows indicating the characteristic fields $H_1$, $H_2$ and $H_{c2}$. For comparison, the magnetic field scale for $H\parallel ab$ is also displayed in the inset.
}
\end{figure*}

The signatures of unusual phase transitions below $H_{c2}^{ab}$ and below $T^* \approx$ 2 K are observed in the field-dependent torque magnetometry $\tau(H)$. The torque magnetometry is known to be extremely sensitive to the small changes in the magnetic susceptibility and usually exhibits a large magnetic hysteresis due to vortex pinning in superconductors~\cite{torque,torque2} (Figs. 3a and 3c). This produces a typical saw-tow shaped curve of the field derivative $\tau(H)$, $d\tau/dH$, for up- and down-sweeps of magnetic fields. In FeSe, however, we observed two additional anomalies at $H_1 \sim$ 15 T and $H_2 \sim$ 22 T, as indicated by arrows in Figs. 3a and 3b, well below the irreversible field $H_{\rm irr}$ at which the hysteresis in $\tau(H)$ and $d\tau(H)/dH$ starts to develop. The transition fields $H_1$ and $H_2$ follow the distinct dependence on temperature or field angle from  those of $H_{\rm irr}$ $\approx$ $H_{c2}$. Upon increasing temperature or the tilting angle of the magnetic field ($\theta$) from $H\parallel ab$ (Figs. 3a and 3c), $H_{\rm irr}$ systematically decreases, similar to $H_{c2}$ (Figs. 2a and 2b). In contrast, the anomalies are pronounced only at low temperatures below $T^*$ and near $H \parallel ab$ with $\theta \leq$ 15$^{\circ}$. Also the transition fields $H_1$ and $H_2$ remain almost the same with variation of temperature or field angle. 

These observations allow us to rule out the vortex-related phenomena as a possible origin for the anomalies at $H_1$ and $H_2$.
It has been well known that the vortex phase transitions, such as melting or modification of the vortex structure and the peak effect, usually share the similar dependence on temperature and field orientation with the upper critical field $H_{c2}$~\cite{vortex_BSCCO,vortex_MgB2}, particularly in highly anisotropic superconductors~\cite{vortex_BSCCO,vortex_MgB2}, like FeSe~\cite{FeSevortex}. In contrast, the transition fields observed in FeSe are nearly independent of field angle. Instead, this is one of the characteristic features of the field-induced exotic phases in $e.g.$ $\kappa$-(BEDT-TTF)$_2$Cu(NCS)$_2$ and CeCoIn$_5$~\cite{ffloreview,organicreview,kenzelmannreview}.
Furthermore, we found that the anomalies at $H_1$ and $H_2$ are very sensitive to disorder, whose amount is quantified by the residual resistivity ratio, RRR = $\rho(300  {\rm K}$)/$\rho(T_c)$. As shown in Supplementary Fig. S5, the anomalies at $H_1$ and $H_2$ in $d\tau(H)/dH$ are present for the crystals with a large RRR $\geq$ 35. In the crystal with a small RRR $\approx$ 10, we obtained the typical saw-tow shaped  curves of $d\tau(H)/dH$ without additional anomalies. The strong sensitivity to disorder is also consistent with the field-induced phase transition in the Pauli-limiting superconductors~\cite{ffloreview,organicreview,kenzelmannreview}.

The magnetocaloric effect (MCE) provides further thermodynamic evidence regarding the field-induced phase transitions. Once a superconducting sample is placed in a changing magnetic field, difference between the sample temperature ($T$) and the bath temperature ($T_{\rm bath}$) is determined by the entropy ($S$) change and the specific heat $C_s$ contribution and is described by $(\partial S/\partial H)_T=-(C_s/T)(dT/dH)-\kappa (T-T_{\rm bath})/TH-(1/T)dQ_{\rm loss}/dH$. Here, $\kappa$ is the thermal conductivity between the sample and the environments~\cite{MCE}, and $Q_{\rm loss}$ corresponds to an additional heat due to the irreversible effect. In our case, the sample is weakly coupled to the bath, due to a relatively low $\kappa$, $i.e.$ being close to the adiabatic condition.
Then the field dependent temperature $T(H)$ can be considered as an isentropic curve where the decrease (increase) of temperature implies the increase (decrease) of entropy.

With increasing magnetic field, we observed weak but clear anomalies in $R_T (H)$ (Figs. 3d and 3e), the resistance of a temperature sensor attached to the crystal, which roughly proportional to the inverse of the sample temperature and becomes similar to the shape of the field dependent entropy, $S(H)$. In conventional superconductors the monotonous field dependence of $R_T (H)$ is usually expected with an anomaly only at $H_{c2}$. However in our experiments we observed additional anomalies in both up- and down-sweeps of magnetic fields at similar $H_1$ and $H_2$ (Supplementary Fig. S6). With variation of temperature and field orientation, these MCE anomalies appears only below $T^*$ and near $H\parallel ab$ with $\theta \leq$ 15$^{\circ}$. The transition fields $H_1$ and $H_2$ are nearly independent of temperature and field angle, in good agreement with the $\tau(H)$ results (Figs. 3a and 3c), as summarized in the magnetic phase diagrams as a function of temperature and field-orientation (Figs. 4a and 4b).

\begin{figure*}[t]
\centering
\includegraphics[width=12cm, bb=30 60 520 300]{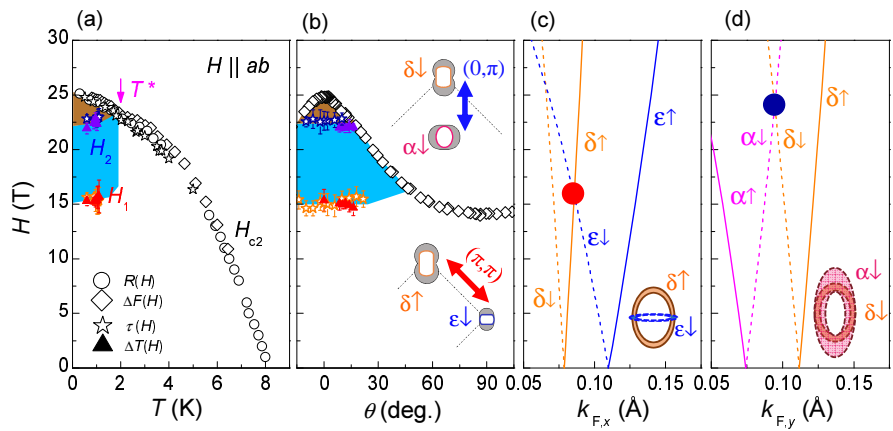}
\caption{\label{fig4}(a,b) Magnetic phase diagram of FeSe for $H \parallel ab$ as a function of temperature (a) and field orientation ($\theta$) with respect to the $ab$ plane (b). The upper critical field $H_{c2}^{ab}(T)$ (open symbols) are determined by resistivity (black open circle), TDO frequency (black open diamond), and torque magnetometry (black open star) measurements, while the field-induced phase transitions at $H_1$ and $H_2$ are obtained by torque magnetometry (colored open star) and the magnetocaloric effect (colored solid triangle). The inset shows schematics of the possible nesting effect via $q=(0,\pi)$ or $(\pi,\pi)$ between the spin-split Fermi surfaces (FSs) with anisotropic superconducting gap (gray shade). Near $H_1$, the nesting effect is expected between $\delta$$\uparrow$ and $\epsilon$$\downarrow$ FSs via a SDW momentum $q=(\pi,\pi)$ (bottom), while the nesting with $q=(0,\pi)$ is expected near $H_2$ between $\alpha$$\downarrow$ and $\delta$$\downarrow$ FSs (top). (c,d) Magnetic field dependent Fermi vectors along two orthogonal directions of $k_x$ (c) and $k_y$ (d) for $\uparrow$ (solid line) and $\downarrow$ (dashed line) spins in the plane of $k_z = 0$. The good matches between the Fermi vectors of $\varepsilon$$\downarrow$- and $\delta$$\uparrow$-FSs along the $k_x$ direction near $H_1$ and also between those of $\delta$$\downarrow$- and $\alpha$$\downarrow$-FSs along $k_y$ direction near $H_2$ are indicated by red and blue circles, respectively. The corresponding nesting conditions between different pairs of the spin-split FSs at $H_1$ and $H_2$ are schematically shown in the insets. The $k_z$ warping of each FS is indicated by different FS cross-section in the planes of $k_z$ =0 (inner line) and $k_z$ = $\pi$ (outer line).}
\end{figure*}
The magnetic field dependent electronic specific heat $C_s/T$ of FeSe also shows consistent results. In Fig. 3f, we plot $C_s/T$, taken at 0.35 K, as a function of the normalized magnetic field $h$ = $H/H_{c2}$ for $H \parallel c$ and $H \parallel ab$. In both cases, $C_s/T$ increases rapidly at low magnetic fields below $h$ $<$ 0.3, which indicates the field-induced closing of smaller SC gaps mostly in the $\delta$- and the $\epsilon$-FSs, consistent with the previous report~\cite{FeSe:Cp}. Thus at higher magnetic fields, the field dependence of $C_s/T$ is dominated by the contribution from the $\alpha$-FS and is expected to follow a nearly linear field dependence below $H_{c2}$ for both $H \parallel c$ and $H \parallel ab$. This would make $C_s(h)/T$ curves collapsed into a single curve, once plotted as a function of the normalized $h$. However in FeSe at higher fields, $C_s(h)/T$ exhibit different behaviour depending on the field orientations. For $H \parallel c$, $C_s(h)/T$ increases almost linearly until it is saturated at $H_{c2}$, as expected. In contrast, for $H \parallel ab$, $C_s(h)/T$ deviates from the linear field dependence at $h$ $\approx$ 0.5 with a slope change, and then recovers its normal state value at $H_{c2}$. The difference $\Delta C_s(h)/T$ = $C_s(h)/T|_{H\parallel ab}$-$C_s(h)/T|_{H\parallel c}$ reveals the anomaly more clearly. The drop of $\Delta C_s(h)/T$ by $\sim$ 0.4 mJ/mol K$^2$ occurs clearly at $h$ $\approx$ 0.5, corresponding to $H_1$ $\sim$ 15 T, while the anomaly $H_2$ $\sim$ 22 T appears to be smeared out near $H_{c2}$.
Combined with anomalies seen in torque and magnetocaloric measurements, the specific heat anomaly confirms the presence of field-induced phases near the in-plane magnetic fields below $H_{c2}^{ab}$.

\section{Discussion}

Now we discuss possible scenarios for the in-plane field-induced phase transitions in FeSe. One candidate is the magnetic Lifshitz transition, involving the field-induced change of the FS topology. This transition has been proposed to occur in FeSe due to the small $E_{\rm F}$~\cite{ptok}, which is however unlikely because the following reasons. Firstly, in order to have a magnetic Lifshitz transition at $H$ $\sim$ 20 T, $e.g.$ in the $\alpha$-FS having the smallest $A$ at $k_z$=0, the $g$-factor of FeSe must be $\sim$5-14, unreasonably large for FeSCs~\cite{gfactor}. Secondly, the transition field for magnetic Lifshitz transition is set by electronic structures and is not directly related to the presence of the superconducting order. It is thus expected to occur even in the normal state, which is not the case in experiments~\cite{ptok}. Experimentally the field-induced phase transitions are only observed in the narrow ranges of temperature and field angle within the superconducting phase, and their phase boundaries are never extended to the normal state (Figs. 4a and 4b).

One possible candidate is then the FFLO phase~\cite{ffloreview,organicreview}. In this case, the Cooper-pair state ($k \uparrow$, $-k+q \downarrow$) is formed with a momentum mismatch ($q$) between the spin-split FSs, inducing a spatially modulated superconducting order with a characteristic length of $q^{-1}$. In multiband superconductors, each FS has its own FFLO instability with a different modulation length $q_i^{-1}$ ($i$, band index), and these instabilities compete with each other, inducing the multiple FFLO orders at different magnetic fields~\cite{FFLO1,FFLO2}. It is however rather unlikely that both transitions observed at $H_1$ and $H_2$ in FeSe are due to the multiple FFLO transitions. Such a multiple FFLO ordering, if exists, is fragile to the interband coupling, because the FFLO instabilities with unequal $q$'s would average the interband pairing energy $\sim$ $\Delta_1 \Delta_2$${\cos}$[($q_1$-$q_2$)$r$] to zero ($\Delta_i$, the superconducting gap at a different band $i$), and strongly reduces $T_c$ under magnetic fields. Therefore, multiple FFLO transitions are allowed only if the interband coupling are drastically suppressed. As discussed below, however, the nesting between spin-split FSs in FeSe are expected to be enhanced near $H_1$ and $H_2$ (Figs. 4c and 4d). Furthermore, we found that the entropy is reduced across $H_1$ and $H_2$ for the up-sweep (Figs. 3d and 3e). 
For typical FFLO transitions, entering to the FFLO phase in the up-sweep usually increases the entropy~\cite{entropy1,entropy2} because the additional quasiparticles are introduced due to the spatially-inhomogeneous superconducting gap.

Another promising candidate is the spin-density-wave (SDW) phase of field-induced quasiparticles. 
As it has been discussed for a $d$-wave superconductor CeCoIn$_5$~\cite{Q_SDW1,Q_SDW2,kenzelmannreview}, the SDW order can be triggered by the nesting effect in the $k$-space near the superconducting gap nodes, where the superconductivity is suppressed by Pauli pair breaking. This nesting effect together with the residual AFM fluctuations is enhanced only in the superconducting state~\cite{Q_SDW1,Q_SDW2}, allowing the coexisting SDW and superconducting phase.
Recent experimental and theoretical studies~\cite{FeSe:QPI,FeSe:ARPES,FeSeGap1,FeSeGap2} revealed that the hole ($\alpha$) and the electron ($\delta$ and $\epsilon$) FSs have the opposite in-plane momentum anisotropy in their superconducting gap as shown in Fig. 1c. Furthermore, two different types of spin fluctuations are dominant in FeSe, the stripe-type spin fluctuation with $q=(\pi,0)$ and the Neel-type one with $q=(\pi,\pi)$~\cite{FeSepipi2}. Therefore, if the incipient spin fluctuations with $q=(\pi,0)$ or $q=(\pi,\pi)$ strongly couple the field-induced quasiparticles near the superconducting gap nodes or minima, the field-induced SDW orders can coexist with the superconducting state and induce the SDW transitions in FeSe. Note that such a SDW order is stable only within the superconducting phase, which is consistent with our phase diagram (Fig. 4a and 4b) and compatible with the absence of FS reconstruction above $H_{c2}$~\cite{FeSe:QO1}.

In order to consider the nesting instability of quasiparticles near the superconducting gap nodes or minima, we estimate the magnetic field dependent Fermi vectors $k_F$ in the planes of $k_z$=0 (Figs. 4c, 4d) and $k_z$=$\pi$ (Supplementary Fig. S9) along $k_x$ and $k_y$ directions. Unlike other superconductors, FeSe has an exceptionally small $E_{\rm F}$, and the Zeeman effect with a typical $g$-factor of 2~\cite{gfactor} results in the spin-split FSs much different in size. This produces a large spin imbalance (Fig. 1d) and also significantly affects the nesting condition. We found that the nesting via $q=(\pi,\pi)$ nicely matches the $\epsilon$-FS of $\uparrow$ spins with the $\delta$-FS of $\downarrow$ spins along the $k_x$ direction at H $\sim$ 16 T (Fig. 4c), in good agreement with the observed transition field $H_1$ $\sim$ 15 T. This nesting effect with $q=(\pi,\pi)$ and the resulting interband scattering would suppress the FFLO instability. However, this can couple the field-induced quasiparticles near the nodal superconducting gap regions of the $\delta$ and the $\epsilon$ FSs (the inset of Fig. 4b) through the Neel-type spin fluctuation~\cite{FeSepipi2}, leading to the phase transition at $H_1$. In this case, the entropy is expected to be reduced upon entering the SDW phase in the up-sweep due to decrease of the number of degrees of freedom, in agreement with the experiments (Figs. 3d and 3e). Moreover the specific heat is also expected to be reduced by the SDW gap formation of quasi-particles, again consistent with experiments (Figs. 3f). The observed reduction of $C_s/T$ by $\sim$ 0.4 mJ/mol K$^2$ (Fig. 3f) is $\sim$ 10\% of the total density of states of $\delta$- and $\epsilon$-FSs, indicating that only quasiparticles near the nodal gap region participate the SDW formation.

The higher field phase transition at $H_2$ $\sim$22 T appears to be more intriguing than one at $H_1$. We found another nesting condition via $q=(0,\pi)$ at $H$ $\sim$24 T, close to $H_2$, at which the $\alpha$-FS of $\downarrow$ spins matches well with $\delta$-FS of $\downarrow$ spins along the $k_y$ direction at $k_z$=0 (Fig. 4d) and also along the $k_x$ direction at $k_z$=0 (Supplementary Fig. S9). In this case, however, this nesting condition involves 
the states of $\alpha$-FSs with the superconducting gap maxima 
(the inset of Fig. 4b), which does not favor the field-induced SDW order.
Instead, the FFLO phase due to the intraband superconducting coupling can be a candidate, dominantly in the $\alpha$-FS, which has the largest superconducting gap and is not involved in the possible SDW ordering at $H_1$. Our calculations taking into account the multiband Pauli paramagnetic effect and the FFLO instability, but not the superconducting gap anisotropy, show that the hidden FFLO phase can stabilized at low temperatures (Supplementary Fig. S7), supporting this possibility. We also note that recent thermal conductivity study under the in-plane magnetic field reveal a clear kink at $H_2$, which is attributed to the FFLO phase transition~\cite{FeSe_matsuda}. The detailed nature of the field induced phase transitions, including possible coexistence of the SDW and the FFLO orders hosted by different FSs, remains to be clarified. Nevertheless these observations clearly demonstrate that FeSe offers a unique system, in which field-induced phase transitions occur in the superconducting state.

\section{Conclusion}
 We reported that FeSe undergoes multiple phase transitions under high in-plane magnetic fields below the upper critical fields.
 The extremely small Fermi energy, comparable with the energy scales of the superconducting gap and the field-induced Zeeman effect, is found to be essential to trigger exotic orders with a large spin imbalance, related with either the spin-density-wave or the FFLO phases. These findings add another intriguing aspect to the unique iron-based superconductor FeSe~\cite{FeSe:BCS,FeSeHc2_1,FeSe:BCS2,FeSereview,FeSe:QO1,FeSe:QO2,FeSe:ARPES,FeSe:QPI,FeSe:Cp}, and also pose a challenge to our understanding on the complex interplay between anisotropic superconducting order, incipient magnetic instabilities, and the multi-band effect in the largely spin-imbalanced superconducting systems.


\section*{ACKNOWLEDGMENTS}

The authors thank Y. Matsuda and D.J.Jang for fruitful discussion. We also thank H. G. Kim in Pohang Accelerator Laboratory (PAL) for the technical support. This work was supported by the Institute for Basic Science (IBS) through the Center for Artificial Low Dimensional Electronic Systems (no. IBS-R014-D1) and also by the National Research Foundation of Korea (NRF) through SRC (Grant No. 2018R1A5A6075964) and the Max Planck-POSTECH Center for Complex Phase Materials
(Grant No. 2016K1A4A4A01922028). W.K. acknowledges the support by NRF (No. 2018R1D1A1B07050087, 2018R1A6A1A03025340), and Y.J.J. was supported by NRF (No. 2016R1A2B4016656).

\end{document}